\documentstyle[11pt]{article}

\input epsf
\input{tcilatex}
\begin{document}

\null
\hfill CERN-TH/98-363

\vspace{50pt}

\begin{center}
{\LARGE Quantum Conformal Algebras}
\vskip .3truecm
{\LARGE and Closed
Conformal Field Theory}
\end{center}

\vspace{6pt}

\begin{center}
{\sl Damiano Anselmi}

{\it Theory Group, CERN, Geneva, Switzerland}
\end{center}

\vspace{8pt}

\begin{center}
{\bf Abstract}
\end{center}
We investigate the quantum conformal algebras of
N=2 and N=1 supersymmetric gauge theories.
Phenomena occurring at strong coupling are analysed
using the Nachtmann theorem and very general,
model-independent, arguments.
The results lead us to
introduce a novel class of conformal field theories, 
identified by a
closed quantum conformal algebra. We conjecture that
they are the exact solution to the strongly coupled large-$N_c$ limit
of the open conformal field theories.
We study the basic properties
of closed conformal field theory
and work out the operator product 
expansion of the conserved current multiplet ${\cal T}$.
The OPE structure is uniquely
determined by two central charges, $c$ and $a$. 
The multiplet ${\cal T}$ 
does not contain just the stress-tensor,
but also $R$-currents and
finite mass operators. For this reason, the ratio $c/a$ is different 
from 1. On the other hand,
an open algebra contains an infinite tower of non-conserved currents,
organized in pairs and singlets with respect to 
renormalization mixing. ${\cal T}$ mixes with a second multiplet
${\cal T}^*$ and
the main consequence
is that $c$ and $a$ have different subleading corrections.
The closed algebra simplifies considerably at $c=a$, where it coincides
with the N=4 one.

\vskip 4.5truecm

\noindent CERN-TH/98-363

\noindent November, 1998.

\vfill\eject

Recently \cite{ics,high} we developed 
techniques to study the operator product
expansion of the stress-energy tensor, with the purpose 
of acquiring a deeper knowledge of conformal field theories in four dimensions
and quantum field theories interpolating between pairs of them.
These techniques are similar to those used,
in the context of the deep inelastic 
scattering \cite{muta}, to study the parton-electron scattering
via the light-cone
operator product expansion of the electromagnetic current.
The investigation of the ``graviton-graviton'' scattering, i.e.
the $TT$ OPE, is convenient in a more theoretical context,
to study conformal windows and hopefully the low-energy limit
of quantum field theories in the conformal windows.

Furthermore, an additional ingredient, supersymmetry, 
reveals that a nice algebraic structure \cite{ics}
governs the set of operators generated by
the $TT$ OPE. We called this structure
the {\it quantum conformal algebra} of the theory, since
it is the basic algebraic notion 
identifying a conformal field theory 
in more than two dimensions.
We have considered in detail the maximal supersymmetric 
case in ref. \cite{ics} and in the present paper we extend our 
investigation to N=2 and N=1 theories, with special attention to
the theories with vanishing beta function.

We believe that this interplay 
between physical techniques and more theoretical
issues will be very fruitful for both.

It was observed in \cite{ics} that the relevant features 
of the algebra do not change with the value of the coupling constant.
This was proved using a theorem due to Nachtmann \cite{nach}, found
1973 in the context of the theory of deep inelastic scattering.
Only at special values $g_*$ of the coupling constant
can the algebra change considerably.
One special point is of course
the free-field limit, where infinitely many currents
are conserved. Another remarkable point is the limit
in which the operator product expansion 
closes with a finite set of conserved currents,
which means only the stress-tensor in the N=4 theory, 
but more than the stress
tensor in the N=2 algebra, as we will see.

This special situation, we believe, deserves 
attention {\sl per se}. It is the simplest conformal field theory 
in four dimensions, simpler than free-field theory and yet non-trivial.
It can be viewed as the true 
analogue of two-dimensional conformal field theory.
Because of its simplicity, it is suitable for an algebraic/axiomatic 
investigation. It is expected to be relevant
both physically and mathematically.
For example, in \cite{high} (sect. 4.5) we argued, using the 
AdS/CFT correspondence \cite{malda}, in particular the results of
\cite{klebanov}, that the limit in which the 
$TT$ OPE closes should be the 
strongly coupled large-$N_c$ limit.
In the present paper we argue something similar about 
finite N=2 theories.

The plan of the paper is as follows.
In sections 1 and 2 we study the quantum conformal algebras of the N=2 
vector multiplet and hypermultiplet, respectively. In sections 3 and 4
we combine the two of them into a finite N=2 theory and discuss the 
most important phenomena that take place when the interaction 
is turned on, like renormalization
splitting and renormalization mixing, anomalous dimensions
and so on. In the rest of the paper we argue, using the Nachtmann theorem and
very general arguments, that the algebra closes in the strongly 
coupled large-$N_c$ limit (sect. 5).
We describe various properties of 
closed conformal field theory (sect. 5), compare them to those
of open conformal field theory (sects. 5 and 6), 
give the complete OPE algebra in the N=2 case (section 6)
and discuss aspects of the N=1 closed quantum conformal algebra.

For supersymmetry,
we use the notation of Sohnius \cite{Sohnius}, converted to the Euclidean
framework
via $\delta_{\mu\nu}\rightarrow-%
\delta_{\mu\nu}$, $T^{V,F,S}\rightarrow -T^{V,F,S}$
(these are the vector, spinor and scalar contributions to the stress-tensor), 
$\varepsilon_{\mu\nu\rho%
\sigma} \rightarrow -i\varepsilon_{\mu\nu\rho\sigma}$ and $\gamma_\mu,
\gamma_5\rightarrow -i\gamma_\mu,-i\gamma_5$. Moreover, we multiply $A_i$ 
by a factor $\sqrt{2}$ and use ordinary Majorana spinors $\lambda_i$
instead of symplectic Majorana spinors $\lambda_s^i$
($\lambda_s^i={1\over 2}[(\delta_{ij}-i\varepsilon_{ij})-\gamma_5 
(\delta_{ij}+i\varepsilon_{ij})]\lambda_j$). 
For the current algebra we use the notations of \cite{ics,high}.

\section{Vector multiplet}

We begin our analysis with the N=2 vector multiplet and the
hypermultiplet,
by repeating the steps of \cite{ics}.
The current multiplets have length 2 in spin units, but the important 
point is that {\sl all} of them have length 2. We recall that 
the stress-tensor 
multiplet has length 0 in the N=4 algebra \cite{ics}.
Moreover, there is one multiplet for each spin, even or odd.

The vector, spinor and scalar contributions 
to the currents of the N=2
vector multiplet are
\begin{eqnarray*}
{\cal J}^{V} &=&F_{\mu \alpha }^{+}\overleftrightarrow{\Omega }_{{\rm even}%
}F_{\alpha \nu }^{-}+{\rm impr}.,\quad \quad \quad \quad \quad \quad {\cal A}^{V} =F_{\mu \alpha }^{+}\overleftrightarrow{\Omega }_{{\rm odd}%
}F_{\alpha \nu }^{-}+{\rm impr}., \\
{\cal J}^{F}&=&\frac{1}{2}\bar{%
\lambda}_{i}\gamma _{\mu }\overleftrightarrow{\Omega }_{{\rm odd}}\lambda
_{i}+{\rm impr}.,\quad \quad \quad \quad \quad \quad {\cal A}^{F}=\frac{1}{2}\bar{%
\lambda}_{i}\gamma _{5}\gamma _{\mu }\overleftrightarrow{\Omega }_{{\rm even}%
}\lambda _{i}+{\rm impr}., \\
{\cal J}^{S} &=&M\overleftrightarrow{\Omega }_{{\rm even}%
}M+N\overleftrightarrow{\Omega }_{{\rm even}}N+%
{\rm impr}.,\quad {\cal A}^{S}=-2iM\overleftrightarrow{\Omega }_{{\rm odd}%
}N+{\rm impr}.,
\end{eqnarray*}
where $\overleftrightarrow{\Omega }_{{\rm even/odd}}$ denotes an even/odd
string of derivative operators $\overleftrightarrow{\partial }$ and
``impr.'' stands for the improvement terms \cite{high}.
A simple 
set of basic
rules suffices to determine the operation $\delta _{\zeta
}^{2} $ which relates the currents of a given multiplet and is
a certain combination of
two supersymmetry transformations 
(see \cite{ics} for details).
The result is
\begin{eqnarray*}
{\cal J}_{2s}^{S} &\rightarrow &-2~{\cal A}_{2s+1}^{F}+2~{\cal A}_{2s+1}^{S},
\qquad \qquad
{\cal A}_{2s-1}^{S}\rightarrow -2~{\cal J}%
_{2s}^{F}+2~{\cal J}_{2s}^{S}, \\
{\cal J}_{2s}^{F}&\rightarrow & -8~{\cal A}%
_{2s+1}^{V}+~{\cal A}_{2s+1}^{S},
\qquad \qquad
{\cal A}_{2s-1}^{F}\rightarrow -8~{\cal J}%
_{2s}^{V}+~{\cal J}_{2s}^{S}, \\
{\cal J}_{2s}^{V} &\rightarrow &-2~{\cal A}_{2s+1}^{V}+\frac{1}{4}~{\cal A}%
_{2s+1}^{F},
\qquad \qquad
{\cal A}_{2s-1}^{V} \rightarrow -2~{\cal J}_{2s}^{V}+\frac{1}{4}~{\cal J}%
_{2s}^{F}.
\end{eqnarray*}
As we see, the algebra is more symmetric than the N=4 one \cite{ics}. 
In particular, there is an evident even/odd-spin
symmetry that was missing in \cite{ics}.
We have fixed the normalization of the scalar
axial current ${\cal A}^S$ (absent in N=4) in order to
exhibit this symmetry.
We recall that $T^{V}=-2{\cal J}_2^{V},$ $T^{F}=%
{\cal J}_2^{F}/2$ and $T^{S}=-{\cal J}_2^{S}/4$ are the various contributions
to the stress-tensor.

The list of current multiplets generated by the $TT$ OPE is easily
worked out and reads
\begin{equation}
\begin{tabular}{lllll}
${\cal T} _{0}={1\over 2}{\cal J}_{0}^{S}$ &  
 &    \\ 
${\cal T}_1=-{\cal A}_{1}^{F}+{\cal A}_{1}^{S}$ &   $\Lambda_{1}={1\over 4}{\cal A}_{1}^{F}+{1\over 4}{\cal A}_{1}^{S}$ &    \\ 
${\cal T}_2=8{\cal J}_{2}^{V}-2{\cal J}_{2}^{F}+{\cal J}_{2}^{S}$ &   
$\Lambda_{2}=-2{\cal J}_{2}^{V}-{1\over 2}{\cal J}_{2}^{F}+{3\over 4}{\cal J}_{2}^{S}$ &   $%
\Xi _{2}=\frac{4}{15}{\cal J}_{2}^{V}+\frac{4}{15}{\cal J}_{2}^{F}
+{1\over 5}{\cal J}%
_{2}^{S}$ \\ 
$\Delta_{3}={3\over 7}{\cal A}_{3}^{V}+{15\over 56}{\cal A}_{3}^{F}+
{5\over 28}{\cal A}_{3}^{S}$ &  
$\Lambda_{3}=8{\cal A}_{3}^{V}-2{\cal A}_{3}^{F}+{\cal A}_{3}^{S}$
&   $\Xi _{3}=-{8\over 3}{\cal A}_{3}^{V}-{1\over 3}{\cal A}_{3}^{F}+
{2\over 3}{\cal A}_{3}^{S}$  \\ 
$\Delta _{4}=-{3}{\cal J}_{4}^{V}-{1\over 4}{\cal J}_{4}^{F}+
{5\over 8}{\cal J}_{4}^{S}$  &   $\Upsilon_{4}=\frac{8}{15}{\cal J}_{4}^{V}+\frac{4}{15}
{\cal J}_{4}^{F}+{1\over 6}{\cal J}_{4}^{S} $ &   $%
\Xi _{4}=8{\cal J}_{4}^{V}-2{\cal J}_{4}^{F}+{\cal J}%
_{4}^{S}$ \\ 
$\Delta_5=8{\cal A}_{5}^{V}-2{\cal A}_{5}^{F}+{\cal A}_{5}^{S}$
&  $\Upsilon _{5}=-
\frac{16}{5}{\cal A}_{5}^{V}-{1\over 5}{\cal A}_{5}^{F}+
{3\over 5}{\cal A}_{5}^{S}$& 
$\Omega _{5}=%
\frac{20}{33}{\cal A}_{5}^{V}+{35\over 132}{\cal A}_{5}^{F}+
{7\over 44}{\cal A}_{5}^{S}$ 
 \\ 
$\ldots$ &   $%
\Upsilon_{6}=8{\cal J}_{6}^{V}-2{\cal J}_{6}^{F}+%
{\cal J}_{6}^{S}$&   $\Omega  _{6}=-{10\over 3}{\cal J}_{6}^{V}-{1\over 6}{\cal J}_{6}^{F}+{7\over 12}{\cal J}_{6}^{S}$ \\ 
&   $\ldots$ &  $\Omega _{7}=8{\cal A}_{7}^{V}-2{\cal A}_{7}^{F}+{\cal A}_{7}^{S}$
\end{tabular}
\label{ulla}
\end{equation}
The lowest components of each current multiplet (${\cal T}_2$, $\Lambda_3$, 
$\Xi_4$, $\Delta_5$, $\Upsilon_6$, $\Omega_7$)
have the same form.
The normalization is fixed in such a way that these components have also
the same overall factor.
In \cite{high} we used a different convention:
we fixed the normalization of each current
by demanding that the coefficients of ${\cal A}^{F}$ and ${\cal J}^{S}$
be 1. Here we have to be more precise and 
keep track of the relative factors inside 
current multiplets, since
we need to superpose the vector and matter quantum conformal
algebras in order to obtain the most general 
N=2 structure (see section 3).

{\it Checks}. Scalar odd-spin currents appear in the algebra and 
their two-point functions were not computed in \cite{high}.
We can combine 
orthogonality checks with the indirect derivations 
of the amplitudes of these 
currents.

These currents are necessary to properly diagonalize the multiplet.
For example, 
only the ${\cal A}^S_1$-independent
combination $-{1\over 2}{\cal T}_1+2\Lambda_1=
{\cal A}^F_1$ appears in the OPE,
but the scalar current ${\cal A}^S_1$ orthogonalizes 
${\cal T}_1$ and $\Lambda_1$.
Indeed, the two-point function of the scalar
spin-1 current, easy to compute,
\[
\langle {\cal A}^S_\mu(x)\,{\cal A}^S_\nu(0)\rangle= {4\over 3}N_V \left( {1\over 4\pi^2}\right)^2
\,\pi_{\mu\nu} \left( {1\over |x|^4}\right),
\]
agrees with the orthogonality 
relationship $\langle {\cal T}_1\,\Lambda_1\rangle=0$.
Similarly, 
${\cal T}_2$ and $\Lambda_2$ are orthogonal
and this can be verified with the results of \cite{ics}. 
$\Xi_2$ is then determined
by requiring that it is orthogonal to both 
${\cal T}_2$ and $\Lambda_2$. Note that $\Xi_2$ 
has the same form as $\Xi_2$ in the
N=4 algebra \cite{ics}, apart from a factor
due to the different normalization.

Then $\Lambda_3$ and $\Xi_3$ are determined
via the transformation $\delta _{\zeta
}^{2}$. The two-point function of ${\cal A}^S_3$ is derived by the
orthogonality relationship $\langle \Lambda_3\, \Xi_3\rangle=0$. We obtain
\[
\langle {\cal A}^S_3(x)\,{\cal A}^S_3(0)\rangle={8\over 35}N_V 
\left( {1\over 4\pi^2}\right)^2
\,{\prod}^{(3)} \left( {1\over |x|^4}\right),
\]
while $\langle {\cal A}^F_3\,{\cal A}^F_3\rangle$ and $\langle {\cal A}^V_3\,{\cal A}^V_3\rangle$
can be found in \cite{high}. Then we determine $\Delta_3$ via
the equations $\langle \Delta_3\,\Lambda_3 \rangle=\langle \Delta_3\, \Xi_3\rangle=0$ 
and $\Xi_4$, $\Delta_4$
via the $\delta _{\zeta
}^{2}$ operation. The amplitudes of \cite{high}
suffice to show that $\langle \Xi_4\, \Delta_4\rangle=0$, which is
a non-trivial numerical check of the values.

Finally, once $\Upsilon_4$ is found by solving $\langle \Upsilon_4\, \Xi_4\rangle
=\langle \Upsilon_4\, \Delta_4\rangle=0$,
we extract the two-point function of ${\cal A}^S_5$
via the orthogonality condition of $\Delta_5$ and $\Upsilon_5$, with the result
\[
\langle {\cal A}^S_5(x)\,{\cal A}^S_5(0)\rangle={2^5\over 3^2\cdot 7\cdot 11}N_V 
\left( {1\over 4\pi^2}\right)^2
\,{\prod}^{(5)} \left( {1\over |x|^4}\right).
\]
$\Omega_5$ is determined by the conditions $\langle \Omega_5\, \Upsilon_5\rangle=0$
and $\langle \Omega_5\, \Delta_5\rangle=0$, and so on.

Any current multiplet is 2-spin long and has the form 
\begin{eqnarray}
\Lambda _{s}&=&4~{a_s~{\cal J}_{s}^{V}+b_s~{\cal J}_{s}^{F}+c_s~{\cal J}
_{s}^{S}\over (a_s+8b_s+8c_s)} \qquad \rightarrow \qquad  \nonumber \\
\Lambda_{s+1}&=&{4\over (a_s+8b_s+8c_s)}\left[-2(a_s+4b_s)~{\cal A}_{s+1}^{V}+{1\over 4} (a_s-8c_s)~{\cal A}
_{s+1}^{F}+ (b_s+2c_s)~{\cal A}
_{s+1}^{S}\right] \qquad \rightarrow  \nonumber \\
\Lambda_{s+2}&=&8{\cal J}_{s+2}^{V}- 2~{\cal J}%
_{s+2}^{F}+~{\cal J}_{s+2}^{S}.
\end{eqnarray}
for all $s$ (${\cal J}\leftrightarrow {\cal A}$ when $s$ is odd).

We stress again the most important novelty exhibited
by the N=2 algebra
with respect to the 
N=4 one \cite{ics}: the multiplet of the stress-tensor
is not shorter than the other multiplets;
rather, it contains also a spin-1 current (the $R$-current)
and a spin-0 partner, on which we will have more to say later on.

The theory is not finite in the absence of hypermultiplets.
Nevertheless, it is meaningful to calculate the anomalous dimensions
of the operators to lowest order, since at one-loop order 
around a free-field theory
conformality is formally preserved.
We give here the first few values of the anomalous dimensions
for illustrative purposes.
The procedure for the computation 
is the same as the one of ref. \cite{ics} and 
will be recalled in the next sections.
We find $h_{\cal T}=0$, $h_\Lambda=2N_c {\alpha\over \pi}$ and
$h_\Xi={5\over 2}N_c {\alpha\over \pi}$. These three values obey 
the Nachtmann theorem \cite{nach}, which states that the 
spectrum of anomalous dimensions is a 
positive, increasing and convex function of the spin.
Actually, the Nachtmann theorem applies only to the lowest
anomalous dimension of the even-spin levels.
Nevertheless, it seems that the property is satisfied by all the
spin levels in this particular case. This is not true in the presence
of hypermultiplets, as we will see.

\section{Hypermultiplet}

The structure of the algebra is much simpler for the matter multiplet.
The currents are
\begin{eqnarray*}
{\cal J}^{F}&=&\bar
\psi\gamma _{\mu }\overleftrightarrow{\Omega }_{{\rm odd}}\psi
+{\rm impr}.,\quad \quad \quad \quad \quad \quad 
{\cal A}^{F}=\bar
\psi\gamma _{5}\gamma _{\mu }\overleftrightarrow{\Omega }_{{\rm even}%
}\psi+{\rm impr}., \\
{\cal J}^{S} &=&2\bar A_i\overleftrightarrow{\Omega }_{{\rm even}%
}A_i+{\rm impr}.,
\end{eqnarray*}

The basic operation $\delta^{2}_\zeta$
does not exhibit the even/odd spin symmetry
and is more similar to the N=4 one:
\begin{eqnarray*}
{\cal J}_{2s}^{S} &\rightarrow &-4~{\cal A}_{2s+1}^{F},
\qquad \qquad
\\
{\cal J}_{2s}^{F}&\rightarrow & -2~{\cal A}%
_{2s+1}^{F},
\qquad \qquad
{\cal A}_{2s-1}^{F}\rightarrow -2~{\cal J}%
_{2s}^{F}+~{\cal J}_{2s}^{S}.
\end{eqnarray*}
It gives the following list of multiplets
\begin{equation}
\begin{tabular}{lllll}
${\cal T} _{0}=-{1\over 4}~{\cal J}_{0}^{S}$ &  & 
 &  &  \\ 
${\cal T}_1={\cal A}_{1}^{F}$ &  & &  &  \\ 
${\cal T}_2=-2~{\cal J}_{2}^{F}+{\cal J}_{2}^{S}$ &  & 
  $%
\Xi _{2}=-{1\over 5}~{\cal J}_{2}^{F}-{3\over 20}~{\cal J}%
_{2}^{S}$ &  & \\ 
 &  &
 $\Xi _{3}={\cal A}_{3}^{F}$ \\ 
 &  & $%
\Xi_{4}=-2~{\cal J}_{4}^{F}+{\cal J}%
_{4}^{S}~$ &  & $\Upsilon_{4}=-\frac{2}{9}~%
{\cal J}_{4}^{F}-{5\over 36}~{\cal J}_{4}^{S} $ \\ 
& & & &$\Upsilon _{5}={\cal A}_{5}^{F}$
 \\ 
&  & &  & $%
\Upsilon_{6}=-2~{\cal J}_{6}^{F}+~{\cal J}_{6}^{S}, $%
\end{tabular}
\label{cum}
\end{equation}
determined with the familiar procedure.
We see that no spin-1 scalar current appears and that,
again, the stress-tensor has two partners, the $R$-current
and a mass term. The general form of the current hypermultiplet
is particularly simple:
\[
\Lambda_{2s}=-{a_s~{\cal J}^F_{2s}+b_{s}~{\cal J}^S_{2s}
\over 2(a_s+2b_s)}\quad \rightarrow \quad
\Lambda_{2s+1}={\cal A}^F_{2s+1}\quad \rightarrow \quad
\Lambda_{2(s+1)}=-2~{\cal J}_{2(s+1)}^{F}+~{\cal J}_{2(s+1)}^{S}.
\]
There is no anomalous dimension to compute here,
since the hypermultiplet admits no renormalizable self-coupling.
In the next section we combine vector multiplets 
and hypermultiplets to study in particular the finite N=2 theories.

\section{Combining the two multiplets into a finite N=2 theory}

In this section we work out the quantum conformal 
algebra of finite N=2 supersymmetric theories. 
We consider, as a concrete
example (the structure
is completely general), the theory
with group $SU(N_c)$ and $N_f=2N_c$ hypermultiplets in the fundamental
representation.
The beta-function is just one-loop owing to N=2
supersymmetry. Precisely, it is
proportional to $N_c-{1\over 2}N_f$,
so it vanishes identically for $N_f=2N_c$ \cite{appropriate}.
Combining the free-vector and free-hypermultiplet 
quantum conformal algebras is not as straightforward as it might seem.
The algebra is much richer than the N=4 one and some non-trivial work
is required before singling out its nice properties.

To begin with, the us write down the 
full multiplet ${\cal T}={\cal T}_v+{\cal T}_m$ of the stress-tensor:
\begin{eqnarray}
{\cal T}_0&=&{1\over 2}~{\cal J}_{0v}^S-{1\over 4}{\cal J}_{0m}^S=
{1\over 2}(M^2+N^2-\bar A_iA_i),
\nonumber\\
{\cal T}_1&=&-{\cal A}^F_{1v}+{\cal A}^F_{1m}+{\cal A}^S_{1v}=
-{1\over 2}\bar\lambda_i\gamma_5\gamma_\mu\lambda_i+
\bar\psi\gamma_5\gamma_\mu\psi-2iM\overleftrightarrow{\partial }_\mu N,
\nonumber\\
{\cal T}_2&=&8{\cal J}^V_{2v}-2({\cal J}^F_{2v}+{\cal J}^F_{2m})
+{\cal J}^S_{2v}+{\cal J}^S_{2m}=-4T_{\mu\nu},\nonumber
\label{mul}
\end{eqnarray}
where the additional subscripts $v$ and $m$ refer to the vector and 
matter contributions, respectively (this heavy notation is 
necessary, but fortunately temporary - we write down the explicit 
formulas in order to facilitate the reading).

In general, the full ${\cal T}$-multiplet 
appears in the quantum conformal algebra.
${\cal T}_1$ is the $SU(2)$-invariant $R$-current, and its anomaly
vanishes because it is proportional to the beta-function. 
${\cal T}_0$ is one of the finite mass perturbations \cite{parkes}.
Our picture gives a nice argument for the finiteness of 
such a mass term, which follows directly
from the finiteness of the stress-tensor. 

The next observation is that the ${\cal T}$-multiplet has to be part of a pair
of multiplets having the same position in the algebra. 
The general OPE structure 
of \cite{high} shows that the singularity 
$1/|x|^6$ carries the sum of the squared scalar fields with coefficient 1.
In our case it should be $M^2+N^2+2\bar A_iA_i$
and not just $M^2+N^2-\bar A_iA_i$. On the other hand,
the mass operator
$M^2+N^2+2\bar A_iA_i$ is not
finite and cannot stay with the stress-tensor. 
Therefore it is split into a linear combination of two operators, precisely
${\cal T}_0$ and ${\cal T}_0^*={1\over 2}~
{\cal J}_{0v}^S-{1\over 4}~{\cal J}_{0m}^S={1\over 2}~(
M^2+N^2+\bar A_iA_i)$. These two operators are not 
orthogonal:
they can freely mix under 
renormalization, because
their current multiplets ${\cal T}$ and ${\cal T}^*$
have the same position in the 
algebra.
This means that in the N=2 quantum conformal algebra 
the $I$-degeneracy of \cite{high} survives.
Orthogonalization would be rather awkward, since the number
of components of $M$ and $N$ is proportional to $N_c^2-1$,
while the number of components of $A_i$ is proportional
to $N_f N_c=2N_c^2$. Coefficients of the form
$\sqrt{(N_c^2-1)/N_c^2}$ would appear and the diagonalization
would not survive once the interaction is turned on.
In the presence of mixing, there is no privileged basis
for the two currents, in general.

However, the ${\cal T}\,{\cal T}^*$-pair satisfies a further property, namely
${\cal T}$ and ${\cal T}^*$ do split in the large-$N_c$ limit
(we will present in sects. 5 and 6 an interesting
interpretation of this fact).
We have fixed ${\cal T}_0^*$ by imposing
$\langle {\cal T}_0\, {\cal T}_0^*\rangle=0$ in this limit.
The complete ${\cal T}^*$ 
multiplet is then ${\cal T}^*={\cal T}^*_v-{\cal T}^*_m$:
\begin{eqnarray}
{\cal T}^*_0&=&{1\over 2}~{\cal J}_{0v}^S+{1\over 4}~{\cal J}_{0m}^S=
{1\over 2}(M^2+N^2+\bar A_iA_i)
\nonumber\\
{\cal T}^*_1&=&-{\cal A}^F_{1v}-{\cal A}^F_{1m}+{\cal A}^S_{1v}=
-{1\over 2}\bar\lambda_i\gamma_5\gamma_\mu\lambda_i-
\bar\psi\gamma_5\gamma_\mu\psi-2iM\overleftrightarrow{\partial }_\mu N
\nonumber\\
{\cal T}^*_2&=&8{\cal J}^V_{2v}-2({\cal J}^F_{2v}-{\cal J}^F_{2m})
+{\cal J}^S_{2v}-{\cal J}^S_{2m}=
-4(T_v-T_m),
\nonumber
\label{mul2}
\end{eqnarray}
where $T_v$ and $T_m$ are the vector and matter contributions 
to the stress-tensor.

Now we analyse the spin-1 level of the OPE.
The first observation is that 
the scalar current contribution
${\cal A}^S_{1v}=-2iM\overleftrightarrow{\partial }_\mu N$ 
appears in ${\cal T}_1$ and ${\cal T}_1^*$.
We know that it does not appear in the general free-field
algebra \cite{high}.
Moreover, the relative coefficient of the fermionic contributions
${\cal A}^F_{1v}$ and ${\cal A}^F_{1m}$
(coming from vector multiplets and hypermultiplets) should be 1.
These two conditions cannot be satisfied by taking a
linear combination of ${\cal T}_1$ 
and ${\cal T}_1^*$, so that a new current should appear, precisely
the lowest-spin current of a new multiplet.
This is the multiplet $\Lambda$ of (\ref{ulla}), which is orthogonal
to both ${\cal T}$ and ${\cal T}^*$, and therefore unaffected by the 
hypermultiplets (but only in the free-field limit - see below).
The scalar current $-2iM\overleftrightarrow{\partial }_\mu N$
is required to properly orthogonalize the multiplets,
as it happens in the spin-0 case.

Some effects appear just when the interaction
is turned on:
the scalar current ${\cal A}^S$, which 
cancels out at the level of the free-field algebra,
appears at non-vanishing $g$.
The current $\Lambda$ does not
depend on the hypermultiplets
at the free-field level, but receives hypermultiplet contributions
at non-vanishing $g$. The procedure
for determining the currents at the interacting level is worked out
in \cite{ics}.
In particular, after covariantizing the derivatives we have to take the
traces out. In the construction of $\Lambda$, such traces are proportional
to the vector multiplet field equations, which receive contributions
from the hypermultiplets at $g\neq 0$.

At the spin-2 level of the OPE, the situation is similar to the
spin-0 one. The basic
formulas for the squares are 
\begin{eqnarray*}
\langle {\cal J}^V_{2v}\,{\cal J}^V_{2v}\rangle&=&1/20\, N_V, 
\quad\quad\langle {\cal J}^F_{2v}\,{\cal J}^F_{2v}\rangle=2/5\, N_V,
\quad\quad\langle {\cal J}^S_{2v}\,{\cal J}^S_{2v}\rangle=8/15\, N_V,\\
\langle {\cal J}^F_{2m}\,{\cal J}^F_{2m}\rangle&=&4/5\, N_c^2,
\quad\quad\langle {\cal J}^S_{2m}\,{\cal J}^S_{2m}\rangle=32/15\, N_c^2,
\end{eqnarray*}
factorizing out the common factor
$1/(4\pi^2)^2\,{\prod}^{(2)}(1/|x|^4)$.
Three spin-2 operators come from the previous multiplets, ${\cal T}_2$, 
${\cal T}^*_2$ 
and $\Lambda_2$, and two new operators appear, $\Xi_2$ and $\Xi_2^*$.
These two mix under renormalization and do not split in the large-$N_c$ limit
(see next section).
They have the form
\[
\frac{4}{15}~{\cal J}_{2v}^{V}+\frac{4}{15}~{\cal J}_{2v}^{F}
+{1\over 5}~{\cal J}%
_{2v}^{S}+\alpha_\Xi
\left(2~{\cal J}_{2m}^{F}+{3\over 2}~{\cal J}%
_{2m}^{S}\right)=\Xi_{2v}+\alpha_\Xi \Xi_{2m}.
\]
We call $\alpha_\Xi$ the coefficient for 
$\Xi_2$ and $\alpha_\Xi^*$ the one for 
$\Xi_2^*$.
In order to proceed with the study of the 
quantum conformal algebra, it is not necessary to fix
both $\alpha_\Xi$ and $
\alpha_\Xi^*$,
and we can treat any degenerate pair, such as $\Xi_2$ and $\Xi_2^*$,
as a whole.

Summarizing, the result is that the final algebra contains the multiplets
\begin{eqnarray}
{\cal T}&=&{\cal T}_{v}+\alpha_{\cal T} \,
{\cal T}_{m},\quad\quad\quad\quad \, \,
\Lambda \,\,= \,\,\Lambda_{v},\nonumber\\
\Xi&=&\Xi_{v}+\alpha_\Xi \, \Xi_{m},\quad\quad\quad\quad \,
\Delta  \,\,= \, \,\Delta_{v},\nonumber\\
\Upsilon&=&\Upsilon_{v}+\alpha_\Upsilon \,\Upsilon_{m},\quad\quad\quad\quad
\Omega  \,\,=  \,\,\Omega_{v},\nonumber
\end{eqnarray}
and so on. We have $\alpha_{\cal T}=-\alpha_{\cal T}^*=1$,
while $\alpha_{\Xi}$ and 
$\alpha_{\Upsilon}$ are undetermined. Fixing them by diagonalizing the
matrix of two-point functions is possible to the lowest order (and 
in the next section we use this property to present the results
of our computations),
but in general it is not meaningful to all orders.

\section{Anomalous dimensions and degenerate pairs}

In this section we discuss the currents at non-vanishing $g$, 
compute their anomalous dimensions and study the 
degenerate multiplets.

We start with the spin-1 currents ${\cal T}_1$, ${\cal T}_1^*$
and $\Lambda_1$, which we call $\Sigma^i_\mu$, $i=1,2,3$, respectively.
The currents are easily defined at non-zero coupling $g$ by covariantizing the
derivative appearing in ${\cal A}^S_1$, i.e. ${\cal A}^S_1
\rightarrow -2i M \overleftrightarrow{D}_\mu N$.

The matrix of two-point functions has the form (see for example
\cite{ccfis})
\begin{equation}
\langle \Sigma^i_\mu(x)\,\Sigma_\nu^j(0)\rangle={1\over (|x|\mu)^{h_{ik}(g^2)}}\,\,
\pi_{\mu\nu}\left({c^{(1)}_{kl}(g^2)\over |x|^4}\right){1\over (|x|\mu)^
{h_{jl}(g^2)}}\,\left({1\over 4\pi^2}\right)^2.
\label{ij}
\end{equation}
To calculate the lowest order
of the matrix $h_{ij}(g^2)$ of anomalous dimensions
it is sufficient to take the zeroth-order $c^{(1)}_{ij}(0)$
of the matrix 
of central charges $c^{(1)}_{ij}(g^2)$ (see \cite{high}). 
We have, at finite $N_c$,
\begin{equation}
c^{(1)}_{ij}(0)={8\over 3}\left(\matrix{
2N_c^2-1 & -1 & 0 \cr
-1 & 2N_c^2-1 & 0 \cr
0 & 0 & {1\over 16}N_V
}\right),
\label{c1}
\end{equation}
which becomes diagonal only in the large-$N_c$ limit. Now, from 
(\ref{ij}) we can compute the matrix of divergences
\begin{equation}
\langle \partial\Sigma^i(x)\,\partial\Sigma^j(0)\rangle=
-{3\over \pi^4}{(ch^t+hc)_{ij}\over |x|^8}.
\end{equation}

Calling $a$ the matrix $ch^t+hc$, the explicit computation gives $a=
N_cN_V\, {\alpha\over \pi}\,{\rm diag}
(0,64/3,1)$, whence we obtain
\begin{equation}
h=\left(
\matrix{
0 & {1\over N_c} & 0 \cr
0 & 2N_c-{1\over N_c} & 0 \cr
0 & 0 & 3N_c
}\right)\,{\alpha\over \pi}.
\label{h1}
\end{equation}
This matrix is in general triangular, with entries $(i,3)$ 
and $(3,i)$ equal to zero,
since the current multiplets ${\cal T}$ and $\Lambda$
are orthogonal. Moreover, the entry $h_{11}$ is zero
because of the finiteness of the stress-tensor.
Finally, we observe that the off-diagonal element is suppressed in the 
large-$N_c$ limit, as we expected, and that in this limit
the anomalous dimension of ${\cal T}^*$ becomes
$h_{{\cal T}^*}=2N_c\,{\alpha\over \pi}< h_\Lambda=3N_c\,{\alpha\over \pi}$.

The diagonal form of the pair $({\cal T},{\cal T}^*)$ is given
by $({\cal T}^\prime,{\cal T}^{*\prime})=H\,({\cal T},{\cal T}^*)$
with
\[
H=\left(\matrix{1 & 0 \cr
{1\over 2 N_c^2} & 1-{1\over 2 N_c^2}}\right).
\] 
One finds
$h^\prime={\rm diag}(0,h^*)$ with $h^*={\alpha\over \pi}
\left(2N_c-{1\over N_c}\right)$.

Now we study the spin-2 level of the OPE. A new degenerate pair
$\{\Xi,\Xi^*\}$ appears and therefore we have five currents ${\cal J}^{(i)}_2$,
$i=1,\ldots 5$,
organized into two degenerate pairs and a ``singlet''. The matrix
$c^{(2)}$ of central charges, defined by
\[
\langle {\cal J}^{(i)}_{\mu\nu}(x)\,{\cal J}^{(j)}_{\rho\sigma}(0)\rangle= 
{\frac{1}{%
60}}{1\over (|x|\mu)^{h_{ik}}} 
{\prod}^{(2)}\left({{c^{(2)}_{kl}}\over |x|^4}\right)
{1\over (|x|\mu)^{h_{ik}}}\left({1\over 4\pi^2}\right)^2,
\]
is block-diagonal, $c^{(2)}={\rm diag}(120\,c^{(1)}_{\cal T}, 36 N_V,
c^{(2)}_\Xi)$,
where the first two blocks are proportional
to the spin-1 blocks, see formula
(\ref{c1}). The third block reads
\[
c^{(2)}_\Xi={16\over 5}\left(\matrix{
N_V+{3\over 2}~\alpha_\Xi^2 N_c^2 
    & N_V+{3\over 2}~\alpha_\Xi\alpha_\Xi^{*} N_c^2 \cr
N_V+{3\over 2}~\alpha_\Xi\alpha_\Xi^{*} N_c^2 
     & N_V+{3\over 2}~\alpha_\Xi^{*2} N_c^2 
}\right).
\]
The matrix of divergences is
\begin{equation}
\langle \partial_\mu{\cal J}^{(i)}_{\mu\nu}(x)\, \partial_\rho{\cal J}%
^{(j)}_{\rho\sigma}(0)\rangle= {\frac{3}{4\pi^4}}(c^{(2)}h_2^t+h_2 
c^{(2)})_{ij}\,{\frac{%
{\cal I}_{\nu\sigma}(x)}{|x|^{10}}}. 
\label{ty}
\end{equation}
Again, the matrix $a^{(2)}=c^{(2)}h_2^t+h_2 c^{(2)}$ is block
diagonal and the first two blocks coincide with those of
the corresponding spin-1 matrix. This
correctly reproduces the known anomalous
dimension of ${\cal T}$, ${\cal T}^*$ and $\Lambda$.
Instead, the $\Xi$-block reads
\[
a^{(2)}_\Xi={16\over 5}\left(\matrix{
7+{11\over 2}~\alpha_\Xi^2-2~\alpha_\Xi
  & 7+{11\over 2}~\alpha_\Xi\alpha_\Xi^*-\alpha_\Xi-\alpha_\Xi^* \cr
7+{11\over 2}~\alpha_\Xi\alpha_\Xi^*-\alpha_\Xi-\alpha_\Xi^*
  &7+{11\over 2}~\alpha_\Xi^{*2}-2~\alpha_\Xi^*
}\right)N_cN_V{\alpha\over \pi}.
\]
The calculation that we have performed 
is not sufficient to completely determine the matrix $h$, since 
$h$ is not symmetric. However, at the lowest order, $a^{(2)}$ 
is sufficient for our purpose.
In particular, let us diagonalize $c^{(2)}$ and $a^{(2)}$ in
the large-$N_c$ limit. We have $\alpha,\alpha^*=
{1\over 3}(5\pm \sqrt{31})$ and $h=N_c\, {\alpha\over \pi}\,
{\rm diag}(1.7,3.6)$.
It appears that the entire pair acquires an anomalous dimension and
moves away.

We conclude that the issue of splitting the paired currents
in the large-$N_c$ limit is irrelevant to this case.
What is important is that the two currents 
move together to infinity.
The other pairs of the quantum conformal algebra 
($\Xi$, $\Upsilon$, etc.) exhibit a similar behaviour 
and only the pair ${\cal T}$ is special.

The analysis of the present section
could proceed to the other multiplets
and multiplet pairs that appear in the algebra. However, the description
that we have given so far is sufficient to understand the general
properties of the algebra and proceed. 

We now comment on the validity of the
Nachtmann theorem \cite{nach}, which states that the minimal anomalous 
dimension
$h_{2s}$ of the currents of the
even spin-$2s$ level is a positive, increasing and
convex function of $s$. We have
$h_2=0$ and $h_4={1.7}N_c{\alpha\over \pi}$.
Moreover, $h_\Lambda=3N_c{\alpha\over \pi}>h_4$
and $h_{{\cal T}^*}\sim 2 N_c
{\alpha\over \pi}>h_4$. There is no contradiction with the Nachtmann
theorem, which is restricted to the minimal even-spin values of 
the anomalous dimensions. Nevertheless, it is worth noting that 
the nice regular behaviour predicted by this theorem
cannot be extended in general to the full spectrum of anomalous dimensions.
In particular an odd-spin value $h_{2s+1}$
can be greater than the even-spin value $h_{2(s+1)}$.

Although the spectrum is less regular than in the N=4 case, the
irregularity that we are remarking works in the sense of making certain
anomalous dimensions greater than would be expected.
This will still allow us to argue that 
all anomalous dimensions that are non-zero 
in perturbation theory become infinite in the strongly coupled 
large-$N_c$ limit. In the rest
of the paper we discuss this
prediction and present various consistency checks of it.

Other operators
appear in the quantum conformal algebra besides those that
we have discussed in detail\footnotemark\footnotetext{I am grateful
to S. Ferrara for clarifying discussions of this point.}. 
They can be grouped into 
three classes:

i) symmetric operators with a non-vanishing
free-field limit; these are the ones that we have discussed;

ii) non-symmetric operators with a non-vanishing
free-field limit; these are not completely
symmetric in their indices;

iii) operators with a vanishing
free-field limit; these are turned on by the interaction.

Operators of classes ii) and iii) can often be derived from
those of class i) by using supersymmetry. 
This is the case, for example,
of the N=4 quantum conformal algebra \cite{adrianopoli}. 
The anomalous dimensions are of course
the same as those of their class i)-partners, so that
our discussion covers them and the conclusions that 
we derive are unaffected.

\section{Closed conformal field theory}

The multiplet of the stress-tensor is the most important 
multiplet of the algebra.
Since all of its components are conserved, it will survive 
at arbitrary $g$ and in particular
in the large-$N_c$ limit. The OPE algebra generated by this multiplet
is in general not closed, but it might be closed in some
special cases.

We can classify conformal field theory
into two classes:

i) {\it open} conformal field theory, when the quantum conformal algebra
contains an infinite number of (generically non-conserved) currents;

ii) {\it closed} conformal field theory, when the quantum conformal algebra
closes within a finite set of conserved currents.

This section is devoted to a study of this classification.

We conjecture that closed conformal field theory is
the boundary of the set of open conformal field theories.
Roughly, one can think of a ball centred in
free-field theory. The boundary sphere is the set of closed
theories. The bulk is the set of open theories.
As a radius $r$ one can take the value of the minimal 
non-vanishing anomalous dimension. 
In the N=4 theory, $r$ is the anomalous dimension
of the Konishi multiplet, while in the N=2 
finite theories $r$ is the minimal eigenvalue
of the matrix of anomalous dimensions of 
the $(\Xi,\Xi^*)$-pair. The theory is free for $r=0$,
open for $r<0<\infty$
and closed for $r=\infty$. The function $r=r(g^2 N_c)$ can be taken
as the true coupling constant of the theory instead of $g^2 N_c$.

The Nachtmann theorem is completely general
(a consequence of unitarity and dispersion relations)
and does not make any use of supersymmetry, holomorphy,
chirality or whatsoever,
which would restrict its range  of validity. 
It states in particular
that if $r=0$, then $h_{2s}=0$ $\forall s>0$, 
and if $r=\infty$, then $h_{2s}=\infty$ $\forall s>0$.
The considerations of the previous section, in particular the regularity
of the spectrum of critical exponents,
suggest that in the former
case all current multiplets are conserved and in the latter case all of them
have infinite anomalous dimensions and
decouple from the OPE.
Very precise properties of the strongly coupled limit of the theory
can be inferred from this.

It was pointed out in \cite{high}, using the 
AdS/CFT correspondence \cite{malda}, in particular the results of
\cite{klebanov}, that the $TT$ OPE should close in the strongly coupled
(which means at large 't Hooft coupling, $g^2 N_c\gg 1$) large-$N_c$ 
limit of N=4 supersymmetric theory. In the weakly coupled limit, 
the anomalous dimensions of the various
non-conserved multiplets are non-zero and
$r\sim g^2 N_c$.
The results of \cite{klebanov} 
suggest that in the vicinity of the boundary sphere,
the anomalous dimension of the Konishi operator
changes to $r\sim (g^2 N_c)^{1\over 4}$,
but still tends to infinity.
The Nachtmann theorem then implies that 
all the anomalous dimensions of the non-conserved operators
tend to infinity.

It is reasonable to expect a similar behaviour in the case of
N=2 finite theories (to which the
AdS/CFT correspondence does {\it not} apply, in general).
We expect that in the strongly coupled large-$N_c$ limit 
the OPE closes just with the currents
(\ref{mul}) \footnotemark\footnotetext{We can
call ${\cal T}_0$ the ``spin-0 current'', with some
abuse of language.}. This appears to be the correct generalization of the 
property exhibited in the N=4 case. Therefore we conjecture that

$\cdot$ {\it closed conformal field theory is the exact solution
to the strongly coupled large-$N_c$ limit of open conformal field
theory.}

The weakly coupled behaviour studied in the previous sections
is consistent with this picture.
We have observed that ${\cal T}$
and ${\cal T}^*$ split in the large-$N_c$ limit
already at weak coupling. This suggests that ${\cal T}^*$
moves away from ${\cal T}$. 
Moreover, 
this splitting does not take place in the
other multiplet-pairs ($\Xi$, $\Upsilon$ and so on)
that mix under renormalization:
this means that they remain paired 
and each pair moves to infinity, without leaving any remnant.
 
Secondly, we claim that

$\cdot$ {\it a closed quantum conformal algebra 
determines uniquely the associated closed
conformal field theory\footnotemark
\footnotetext{To our present knowledge, the stronger version of this statement,
i.e. its extension to open algebras, might hold also.
However, this is a more difficult
problem to study.}.}

A similar property holds in two-dimensional conformal field theory
and indeed we are asserting 
that closed conformal field theory is the proper
higher-dimensional analogue of two-dimensional
conformal field theory. Thirdly, we show in the next section that

$\cdot$ {\it
a closed algebra is determined uniquely by two central charges: $c$ and $a$.}

%

The two central charges, called $c$ and $a$ in ref.
\cite{noi} take different 
values in the N=2 algebra, precisely\footnotemark
\footnotetext{We use the normalization of \cite{noi}.}:
\begin{equation}
c={1\over 6}(2N_c^2-1),~~~~~~~~~~~~~~~~~~~
a={1\over 24}(7N_c^2-5),
\label{ca}
\end{equation}
and equal values in the 
N=4 algebra, $c=a={1\over 4}(N_c^2-1)$ if the gauge group is 
$SU(N_c)$.
We recall that the values of $c$ and $a$
are independent of the coupling
constant $g$ \cite{noi} if the theory is finite.

When N=2, the difference between $c$ and $a$
persists in the large-$N_c$ limit
(where $c/a\sim 8/7$),
both strongly and weakly coupled.
The presence of more partners in the current multiplet
of the stress-tensor (precisely 
${\cal T}_0$ and ${\cal T}_1$) is 
related to the ratio ${c\over a}\neq 1$ in N=2 theories, something which we 
will describe better in the next section.
This is a remarkable
difference between closed conformal 
field theory in four dimensions
and conformal field theory in two dimensions,
two types of theories that otherwise
have several properties in common and can be studied in parallel.

Let us now consider N=1 (and non-supersymmetric)
theories. The multiplet of the stress-tensor 
will not contain spin-0 partners, in general, but just the $R$-current.
The above considerations stop at the spin-2 and 1 levels
of the OPE, but the procedure to
determine the closed algebra is the same. What is more subtle
is to identify the physical situation that the closed limit should describe.

In supersymmetric QCD with $G=SU(N_c)$ and $N_f$ quarks and
antiquarks in the fundamental representation, the conformal 
window is the interval
$3/2\, N_c < N_f < 3\, N_c$.
In the limit where both $N_c$ and $N_f$ are large,
but the ratio $N_c/N_f$ is fixed and 
arbitrary in this window, the $TT$ OPE does not close \cite{noialtri}
and $r$ is bounded by
\cite{kogan}
\[
r \sim g^2 N_c < 8\pi^2,
\]
a relationship that assures positivity of the denominator
of the NSVZ exact beta-function \cite{nsvz}.
Therefore the closed limit $r\rightarrow \infty$
presumably does not exist in the conformal window
(it is still possible, but improbable, 
to have $r=\infty$ for some 
finite value of $g^2 N_c$).
The absence of a closed limit 
could be related to the non-integer (rational) 
values of $24 c$ and $48 a$ 
\footnotemark\footnotetext{In our normalization 
the free-field values of $c$ and $a$ are \cite{noi}
$c={1\over 24}(3N_v+N_\chi)$, $a={1\over 48}(9N_v+N_\chi)$,
where $N_v$ and $N_\chi$ are the numbers of vector and chiral multiplets.},
which indeed depend on $N_c/N_f$. 
In the low-energy
limit  we have 
the formulas \cite{noi}
\begin{equation}
c={1\over 16}\left(7N_c^2-2-9{N_c^4\over N_f^2}
\right),~~~~~~~~~~~~~~~~
a={3\over 16}\left(2N_c^2-1-3{N_c^4\over N_f^2}
\right).
\end{equation}
The ${N_c^4\over N_f^2}$-contributions to $c$ and $a$ are not subleading
in the large-$N_c$ limit.
Presumably, an open algebra 
is necessary to produce non-integer values
of $c$ and $a$. This problem deserves
further study and is currently under investigation.

In conclusion, our picture of the moduli space of 
conformal field theory as a ball centred in the origin and with
closed conformal field theory as a boundary
works properly when supersymmetry is 
at least N=2 or, more generally, when
the conformal field theory belongs
to a one-parameter family of conformal field theories
with a point at infinity,
the parameter in question being the radius $r$ of the ball or,
equivalently, the coupling constant $g^2 N_c$.
N=1 finite families of this type are for example
those studied in refs. \cite{lucchesi}.

\section{OPE structure of closed conformal field theory.}

The basic rule to determine the 
quantum conformal algebra of closed conformal field theory
is as follows.
One first studies the free-field OPE of an open conformal
field theory and organizes the currents into orthogonal
and mixing multiplets.
Secondly, one computes the anomalous
dimensions of the operators to the lowest orders 
in the perturbative expansion. 
Finally, one
drops all the currents with
a non-vanishing anomalous dimension.

More generically, one can postulate a set
of spin-0, 1 and 2 currents, that we call
${\cal T}_{0,1,2}$, and study the most general
OPE algebra consistent with closure and unitary.  

The closed limit of the N=4 quantum conformal algebra is very simple
and actually the formula of $\tilde {\rm SP}_{\mu\nu,\rho\sigma;\alpha\beta}$
that was given in \cite{ics}, toghether with the value of 
the central charge $c=a$
encode the full closed N=4 algebra\footnotemark\footnotetext{This 
situation should be described also by the
formalism of N=4 superfields \cite{howe}. Claimed by
its authors to
hold generically in N=4 supersymmetric Yang-Mills theory, 
this formally can actually be correct only in the closed limit
of theories with $c=a+{\cal O}(1)$.}.

We study here the closed N=2 quantum
conformal algebra for generic $c$ and $a$.
The case with gauge group $SU(N_c)$
and $2N_c$ hypermultiplets in the fundamental
representation is $c/a=8/7$.

We have
\begin{eqnarray}
{\cal T}_0(x)\,{\cal T}_0(y)&=&{3\over 8\pi^4}\, 
{c\over |z|^4}-
{1\over 4\pi^2}\left(3-4{a\over c}\right){1\over |z|^2}\,{\cal T}_0(w)
+\,{\rm descendants}+{\rm regular \, terms},
\nonumber\\
{\cal T}_{\mu}(x)\,{\cal T}_{\nu}(y)&=&{c\over \pi^4}\,\pi_{\mu\nu}
\left({1\over |z|^4}\right)-
{4\over 3\pi^2}\left(1-2{a\over c}\right){\cal T}_0(w)\,\pi_{\mu\nu}\left({1\over |z|^2}\right)
\nonumber\\&&
-{1\over \pi^2}\left(1-{a\over c}\right) {\cal T}_\alpha(w)\,
\varepsilon _{\mu \nu \alpha \beta }~\partial _{\beta
}~\left( \frac{1}{|z|^{2}}\right)
\nonumber\\&&
+{1\over 960\pi^2}{\cal T}_{\alpha\beta}(w)
{\prod}^{(1,1;2)}_{\mu,\nu;\alpha\beta}\left(|z|^2 \ln |z|^2M^2\right)
+\,{\rm descendants}+{\rm regular \, terms},\nonumber\\
{\cal T}_{\mu\nu}(x)\,{\cal T}_{\rho\sigma}(y)&=&
{2\, c\over \pi^4}\,{\prod}^{(2)}_{\mu\nu,\rho\sigma}
\left({1\over |z|^4}\right)-{16
\over 3\pi^2}\left(1-{a\over c}\right)
{\cal T}_0(w)\,{\prod}^{(2)}_{\mu\nu,\rho\sigma}
\left({1\over |z|^2}\right)+\nonumber\\&& 
-\frac{4}{\pi ^{2}}\left(1-{a\over c}\right)
{\cal T}_{\alpha }(w)~\sum_{{\rm symm}%
}\varepsilon _{\mu \rho \alpha \beta }~\pi _{\nu \sigma }~\partial _{\beta
}~\left( \frac{1}{|z|^{2}}\right)+\nonumber\\&&
-{1\over \pi^2}\,{\cal T}_{\alpha\beta}(w)\, \tilde 
{\rm SP}_{\mu\nu,\rho\sigma;\alpha\beta}\left({1\over |z|^2}\right)+\,
{\rm descendants}+{\rm regular \, terms},
\label{opppo}
\end{eqnarray}
where $z_\mu=x_\mu-y_\mu$, $w_\mu={1\over 2}(x_\mu+y_\mu)$ and
\begin{eqnarray}
{\prod}^{(1,1;2)}_{\mu,\nu;\alpha\beta}&=&
\left(4+{a\over c}\right){\prod}^{(2)}_{\mu\nu,\alpha\beta}
-{2\over 3}\left(7-5{a\over c}\right)\pi_{\mu\nu}
\partial_\alpha\partial_\beta,\nonumber\\
\tilde{\rm SP}_{\mu\nu,\rho\sigma;\alpha\beta}
\left({1\over |z|^2}\right)&=&
{\rm SP}_{\mu\nu,\rho\sigma;\alpha\beta}\left({1\over |z|^2}\right)
+{1\over 480}\left(102-59{a\over c}\right){\prod}^{(2)}_{\mu\nu,\rho\sigma}
\partial_\alpha\partial_\beta\left(|z|^2 \ln |z|^2M^2\right)\nonumber\\&&
-{1\over 32}\left(12-7{a\over c}\right)
{\prod}^{(3)}_{\mu\nu\alpha,\rho\sigma\beta}\left(|z|^2 \ln |z|^2M^2\right).
\label{uy}
\end{eqnarray}
The numbers appearing in these two space-time invariants are not 
particularly illuminating. 
It might be that the decomposition that we 
are using is not the most elegant one, but for 
the moment we do not have a better one to propose.

The mixed OPE's read:
\begin{eqnarray}
{\cal T}_\mu(x)\,{\cal T}_0(y)&=&{1\over 24\pi^2}\left(1-2{a\over c}\right)
{\cal T}_\nu(w) \, \pi_{\mu\nu}\left(\ln |z|^2M^2\right)+\,
{\rm descendants}+{\rm regular \, terms},\nonumber\\
{\cal T}_{\mu\nu}(x)\, {\cal T}_0(y)&=&{2\over 3\pi^2} {\cal T}_0(w)\,
\pi_{\mu\nu}\left({1\over |z|^2}\right)-{1\over 160\pi^2}
{\cal T}_{\alpha\beta}(w)\left(1-{a\over c}\right)
{\prod}^{(2)}_{\mu\nu,\alpha\beta}(|z|^2
\ln |z|^2 M^2)\nonumber\\&&+\,
{\rm descendants}+{\rm regular \, terms},\nonumber\\
{\cal T}_{\mu\nu}(x)\, {\cal T}_\rho(y)&=&{1\over 2\pi^2}{\cal T}_\alpha(w)
{\prod}^{(2,1;1)}_{\mu\nu,\rho;\alpha}\left(\ln |z|^2M^2\right)
\nonumber\\&&
+{3\over 40 \pi^2}{\cal T}_{\alpha\gamma}(w)\left(1-{a\over c}\right)
\sum_{\rm symm}\varepsilon_{\mu\rho\alpha\beta}
\pi_{\nu\gamma}\partial_\beta(\ln
|z|^2M^2)\nonumber\\&&
+\,{\rm descendants}+{\rm regular \, terms},
\label{mixx}
\end{eqnarray}
where
\begin{eqnarray*}
{\prod}^{(2,1;1)}_{\mu\nu,\rho;\alpha}\left(\ln |z|^2M^2\right)&=&
(\delta_{\alpha\mu}\partial_\nu\partial_\rho+
\delta_{\alpha\nu}\partial_\mu\partial_\rho+2\delta_{\alpha\rho}
\partial_\mu\partial_\nu-\delta_{\mu\rho}\partial_\nu\partial_\alpha
-\delta_{\nu\rho}\partial_\mu\partial_\alpha)\left({1\over |z|^2}\right)\\
&&+{1\over 6}\left(1-2{a\over c}\right){\cal T}_\alpha(w)
{\prod}^{(2)}_{\mu\nu,\rho\alpha}\left(\ln |z|^2 M^2\right).
\end{eqnarray*}

The ${\cal T}_0\,{\cal T}_0$ 
OPE closes by itself, but this fact does not appear
to be particularly meaningful.

We now make some several remarks about the above algebra.

We begin by explaining how
to work out (\ref{opppo})-(\ref{mixx}).
In a generic N=2 finite theory, 
we collect the hypermultiplets into a single representation $R$.
The condition for finiteness is the equality of the Casimirs of $R$ 
and the adjoint representation: $C(G)=C(R)$. 
We have
\[
c={1\over 6}\dim G +{1\over 12} \dim R,~~~~~~~~~~~~~
a={5\over 24}\dim G + {1\over 24} \dim R.
\]

The form of the currents belonging to the
${\cal T}$ multiplet does not depend on the theory
(i.e. on $c$ \& $a$). Instead, the form
of the currents ${\cal T}^*$ does. Let us write
${\cal T}_0^*={1\over 2}{\cal J}^S_{0v}+{\beta\over 4}
{\cal J}^S_{0m}$.
The correlator
$\langle {\cal T}_0\,{\cal T}_0^*\rangle$ is proportional to $\dim G
-{\beta\over 2}\dim R$. By definition, in the closed limit 
$\langle {\cal T}_0\,{\cal T}_0^*\rangle=0$, which gives
\[
\beta=-2 {c-2a\over 5c-4a}.
\]
The scalar operator appearing in the OPE can be decomposed
as
\[
M^2+N^2+2\bar A_i A_i=2 {\beta-2\over \beta+1}{\cal T}_0+
{6\over \beta+1}{\cal T}_0^*.
\]
Dropping ${\cal T}_0^*$, one fixes the coefficient with which
${\cal T}_0$ appears in the $TT$ OPE. It is proportional
to $c-a$, as we see in (\ref{opppo}). The other terms of (\ref{opppo})
and (\ref{mixx}) can be worked out similarly.
 
Let us now analyse the ``critical'' case $c=a$.
Examples of theories with $c=a+{\cal O}(1)$
can easily be constructed. It is sufficient
to have $\dim G= \dim R+{\cal O}(1)$. We do not possess the
complete classification of this case.

At $c=a$ the structure simplifies enormously.
The $TT$ OPE closes just with the stress-tensor, 
$\tilde{\rm SP}_{\mu\nu,\rho\sigma;\alpha\beta}$ reduces to the N=4 
expression \cite{high} and the operators ${\cal T}_{0,1}$ behave as
primary fields with respect to the stress-tensor.

The divergence of the $R$-current (which is simply
${\cal T}_1$ in our notation, up to a numerical factor)
has an anomaly
\begin{equation}
\partial_\mu(\sqrt{g}R^\mu)={1\over 24 \pi^2}(c-a)R_{\mu\nu\rho\sigma}\tilde
R^{\mu\nu\rho\sigma},
\label{ano}
\end{equation}
non-vanishing when $c\neq a$.
The
three-point function $\langle R\,T\,T\rangle$ is not zero, which means that 
the $TT$ OPE does contain some operator mixing with $R_\mu$.
This operator is just $R_\mu$ in the large-$N_c$ limit, 
as we see in (\ref{opppo}),
but {\it both} $R_\mu$ and ${\cal T}_1^*$ at generic $N_c$.
On the other hand, if the theory has $c=a$ and is supersymmetric
(N=1 suffices), then the above anomaly vanishes. Given that the correlator
$\langle R\,T\,T\rangle$ is unique (because there 
is a unique space-time invariant for the
spin-1 level of the OPE, see \cite{high}), we have $\langle R\,T\,T\rangle=0$
in such cases. In conclusion, ${\cal T}_1$ is  
kicked out of the $TT$ OPE algebra 
when $c=a$, even when the number of supersymmetries is less than four. 

The spin-1 
current ${\cal A}^F_{1v}+{\cal A}^F_{1m}$ appearing in the $TT$ OPE
has to be a linear combinarion of $\Lambda_1$ and ${\cal T}_1^*$.
A simple computation shows that
this is possible only for ${\cal T}_1^*=-{\cal A}^F_{1v}-2
{\cal A}^F_{1m}+{\cal A}^S_{1v}$, 
which means $\beta=2$. In turn, this implies that 
${\cal T}_0$ is also out of the $TT$ OPE, as we have just seen.
Therefore the $TT$ OPE closes just with the stress-tensor and
$\tilde {\rm SP}$ coincides precisely with the one given in \cite{ics}:

$\cdot$ {\it the $c=a$ closed algebra 
is unique and coincides with the one of \cite{ics}.}

When $c\neq a$ and supersymmetry is at least N=2
the difference $c-a$ fixes the coefficients of the new terms
in the OPE:

$\cdot$ {\it given $c$ and $a$, there is a unique closed conformal algebra
with N=2 supersymmetry.}

In the large-$N_c$ limit of the N=2 model that
we have studied in section 3,
$c$ and $a$ are ${\cal O}(N_c^2)$ and their ratio is $8/7$. 
At finite $N_c$ (i.e. when the algebra is open)
they receive different
${\cal O}(1)$-order corrections (see (\ref{ca})).
Using our algebra, we can give a very imple explanation of both effects.

The two-point function $\langle TT\rangle$ encodes
only the quantity $c$. $c$ and $a$ are encoded into the three-point
function $\langle TTT\rangle$, 
or the higher-point functions, as it is clear from the
trace anomaly formula
\[
\Theta={1\over 16\pi^2}\left[
c (W_{\mu\nu\rho\sigma})^2-a (\tilde R_{\mu\nu\rho\sigma})^2\right],
\]
and the algebra (\ref{opppo})-(\ref{mixx}),
in particular
the space-time invariant $\tilde {\rm SP}_{\mu\nu,\rho\sigma;\alpha\beta}$
(\ref{uy}).

We can study the three-point function
$\langle T(x)T(y)T(z)\rangle $ by taking the $x\rightarrow y$ 
limit and using the OPE
$T(x)T(y)=\sum_n \hbox{$c_n(x-y)$}{\cal O}_n\left({x+y\over 2}\right)$.
We are lead to consider the two-point functions
$\langle {\cal O}_n\,T\rangle$. Now, we know that the there
are only two operators ${\cal O}_n$
that are not orthogonal to $T$: the stress-tensor itself
{\sl and} ${\cal T}^*$.
${\cal O}_n=T$ produces a contribution
$\langle TT\rangle$, which is again $c$.
This is the contribution to $a$ proportional to $c$.
In the N=2 theories at finite $N_c$
there is a second contribution
from ${\cal O}_n={\cal T}^*$.
Indeed, $\langle {\cal T}^*\, {\cal T}\rangle$ is non-vanishing and
precisely ${\cal O}(1)$, which explains 
the ${\cal O}(1)$-difference between $c$ and 
${8\over 7}a$.
$\langle {\cal T}^*\, {\cal T}\rangle$ is 
not affected by an anomalous
dimension to the second-loop order
(the off-diagonal element of $a^{(2)}_{\cal T}$ vanishes),
which is what we expect, since both $c$ and $a$ are radiatively uncorrected.

The reader might have noted that the operator
${\cal T}_1$ appearing in the ${\cal T}_1{\cal T}_1$ OPE
has a coefficient proportional to $c-a$, 
which means that the $\langle {\cal T}_1{\cal T}_1{\cal T}_1\rangle$
three-point function (which is unique by the usual arguments)
is proportional to $c-a$ and not to $5a-3c$ as in \cite{noi}.
The reason in that our $R$-current is not the same as the one 
that is used in the N=1 context of ref. \cite{noi}: our present $R$-current
is $SU(2)$-invariant, while the one of \cite{noi} is associated with
a $U(1)$ subgroup of $SU(2)$.

Finally, we discuss the embedding of the N=4
open algebra of \cite{ics} into the N=2 open algebra of section 3.
With the terminology ``current multiplet'' we have always referred 
to the subset of components that are generated by the $TT$ OPE.
The quantum conformal algebra 
is embeddable
into a larger set of OPE's, containing
all the supersymmetric partners of the currents that 
we have considered so far.
For example, in the N=4 theory
the $R$-currents (and in particular the object
called ${\cal T}_1$ in the N=2 frame)
are not $SU(4)$ invariant and so they do not appear in the $TT$ OPE,
but they appear in the superpartners of the $TT$ OPE.
From the N=2 point of view, instead, the current
${\cal T}_1$ is $SU(2)$-invariant and 
indeed it does appear in the $TT$ OPE.
Currents that are kicked out of the restricted algebra
are of course always part of the larger web and inside that 
web they ``move''.
Yet what is important is to know
the {\it minimally} closed
algebra.

Let us describe how the Konishi multiplet $\Sigma$ of \cite{ics}
emerges in the N=4 case. The spin-0 current ${\cal T}_0^*$ coincides precisely
with the spin-0 operator $\Sigma_0$ of \cite{ics}.
At the spin-1 level we have, from the N=2 point of view,
two operators: ${\cal T}_1^*$ and $\Lambda_1$ (${\cal T}_1$ having
been kicked out).
The operator $\Sigma_1={\cal A}^F_{1v}+{\cal A}^F_{1m}$
is the linear combination of ${\cal T}_1^*$ and $\Lambda_1$
that does not contain the scalar current ${\cal A}_1^S$, forbidden in the
N=4 algebra by $SU(4)$ invariance. We have $\Sigma_1=-{1\over 2}~{\cal T}_1^*+
2~\Lambda_1$.

Now, the N=2 supersymmetric algebra relates $\Sigma_1$ with 
$-{1\over 2}~{\cal T}_2^*+
2~\Lambda_2$, which, however, is not $\Sigma_2$ and is not $SU(4)$-invariant.
Actually, one has
\begin{equation}
\Sigma_2=-{7\over 20}~{\cal T}_2^*+{7\over 3}~\Lambda_2-2~\Xi_2^*,
\label{sigma2}
\end{equation}
where $\Xi_2^*=\Xi_{2v}+\Xi_{2m}$ is orthogonal 
to the N=4 operator $\Xi_2=5~\Xi_{2v}-{20\over 3}~\Xi_{2m}$.
The N=2 supersymmetric transformation does not generate directly $\Sigma_2$.
$\Sigma_2$ is recovered after a suitable $SU(4)$-invariant
reordering of the currents. The meaning of this fact is that 
the quantum conformal algebra admits various different N=2 (and N=1) 
``fibrations'', depending of the subgroup of $SU(4)$ that one
preserves, and that all of these fibrations are meaningful at arbitrary $g$.

\section{Conclusions}

In this paper we have analysed the quantum conformal algebras of N=2 
supersymmetric theories, focusing in particular on finite theories.
Several novel features arise with respect to the quantum conformal algebra of 
\cite{ics} and each of them has a nice interpretation in the 
context of our approach. 
Known and new properties of supersymmetric theories,
conformal or not, are elegantly grouped by our formal
structure and descend 
from a general and unique mathematical notion.

Stimulated by the results of our analysis,
we have introduced a novel class of conformal field theories
in dimension higher than two, which have a closed quantum 
conformal algebra.
The definition is completely general, valid in any dimension. 
For example, analogous considerations apply to
three dimensional conformal field theory and 
might be relevant for several problems in condensed matter physics.
Closed conformal field theory is nicely tractable from the 
formal/axiomatic point of view. On the other hand, it is interesting
to identify the physical situations or limits of ordinary theories
that it describes. Various considerations suggest that
the closed algebra coincides in general with
the strongly coupled large-$N_c$ limit of an open algebra.
In some cases, like N=1
supersymmetric QCD
as well as ordinary QCD, the identification 
of the closed limit, if any, is more subtle and still unclear.

In a closed conformal field theory the quantities $c$ and $a$ are always
proportional to each other, but the proportionality
factor depends on the theory, i.e. it feels
the structure of the quantum conformal algebra. 
Closed conformal field theory is described uniquely by
these two central charges. We have worked out the 
OPE algebra in detail and related it to known anomaly formulas.
The value of $c$ is given by the two-point functions,
while the value of $a$ is given by the three-point functions,
i.e. by the structure 
of the algebra itself.
There is a critical case, $c=a$, in which the algebra admits a 
closed subalgebra containing only the stress-tensor.

An open conformal field theory presents a richer set of phenomena.
In particular, some operators can mix under renormalization with
the stress-tensor and the main
effect of this mixing is that $c$ and $a$ are not
just proportional. 

We think that 
closed conformal field theory is now ready for a purely algebraic
study, i.e. mode expansion,
classification of unitary representations and so on.
In view of the applications,
we believe that it is worth to proceed in this direction.

\end{document}